\documentclass[conference]{IEEEtran}
\IEEEoverridecommandlockouts
\usepackage{amsmath,amssymb,amsfonts}
\usepackage{graphicx}
\usepackage[dvipsnames]{xcolor}
\usepackage{subcaption}

\def\BibTeX{{\rm B\kern-.05em{\sc i\kern-.025em b}\kern-.08em
        T\kern-.1667em\lower.7ex\hbox{E}\kern-.125emX}}
\begin{document}

\title{Time-Based Roofline for Deep Learning Performance Analysis}

\author{\IEEEauthorblockN{Yunsong Wang, Charlene Yang\\ Steven Farrell, Yan Zhang}
\IEEEauthorblockA{\textit{National Energy Research Scientific Computing Center} \\
\textit{Lawrence Berkeley National Laboratory}\\
Berkeley, CA USA\\
\{yunsongwang, cjyang, sfarrell, yanzhang\}@lbl.gov}
\and
\IEEEauthorblockN{Thorsten Kurth}
\IEEEauthorblockA{\textit{NVIDIA Corporation} \\
Santa Clara, CA USA\\
tkurth@nvidia.com}
\and
\IEEEauthorblockN{Samuel Williams}
\IEEEauthorblockA{\textit{Computational Research Division} \\
\textit{Lawrence Berkeley National Laboratory}\\
Berkeley, CA USA\\
swwilliams@lbl.gov}
}

\maketitle

\begin{abstract}
Deep learning applications are usually very compute-intensive and require a long run time for training and inference. 
This has been tackled by researchers from both hardware and software sides, and in this paper, we propose a Roofline-based approach to performance analysis to facilitate the optimization of these applications. 
This approach is an extension of the Roofline model widely used in traditional high-performance computing applications, and it incorporates both compute/bandwidth complexity and run time in its formulae to provide insights into deep learning-specific characteristics.  
We take two sets of representative kernels, 2D convolution and long short-term memory, to validate and demonstrate the use of this new approach, and investigate how arithmetic intensity, cache locality, auto-tuning, kernel launch overhead, and Tensor Core usage can affect performance. 
Compared to the common ad-hoc approach, this study helps form a more systematic way to analyze code performance and identify optimization opportunities for deep learning applications.

\end{abstract}


\section{Introduction}

Deep learning has gained popularity in recent years due to its high prediction accuracy~\cite{deng2009imagenet,lecun1995convolutional}, and two of the most important neural networks in deep learning are the convolutional neural networks (CNNs) and recurrent neural networks (RNNs).
CNNs can extract both low-level and high-level features in its input data, and have enabled a range of novel applications such as generative adversarial networks (GANs) and variational autoencoders (VAEs)~\cite{goodfellow2014generative}. 
While CNNs are good at dealing with spatial data, RNNs have been widely adopted in temporal data analysis such as time-series prediction and natural language processing (NLP)~\cite{graves2013hybrid,cho2014learning,manning2014stanford}, and one very good example is the long short-term memory (LSTM) neural network.
These neural networks have very deep layers, and their training and inference process is usually a long one due to the large training/test dataset and vast parameter search space~\cite{he2016deep}.


To help understand and improve the performance of these deep learning applications, benchmarking, profiling and performance analysis is crucial.
Several deep learning frameworks and vendors have provided profiling tools to help tackle this problem~\cite{torchprofiler,tfprofiler,dlprof}, and research has been conducted on more detailed analysis of the key computational kernels~\cite{jorda2019performance} or representative workloads~\cite{ren2019performance}.
Unfortunately, most of the performance analysis so far~\cite{Wang_2020_CVPR,Dolhasz_2020_CVPR,Schops_2020_CVPR, shao2017performance,carneiro2018performance} has been focused solely on the run time, and not much insight has been provided on how different networks, algorithms, or frameworks utilize the different hardware capabilities, such as compute capability for different data precisions and memory bandwidth on various cache levels.

\begin{figure}[!htp]
\centering
\includegraphics[width=0.9\columnwidth]{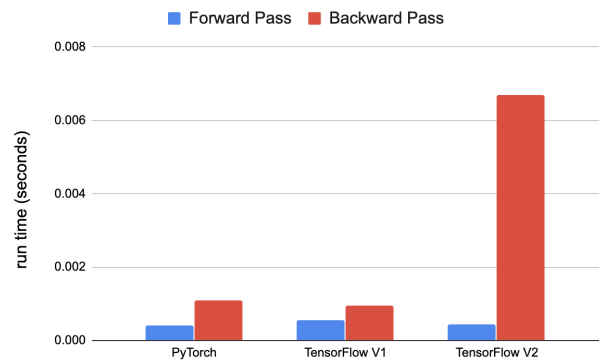}
\caption{Run time of a 2D convolution kernel with 256 filters in half-precision on a V100 GPU. Although TensorFlow~v2's performance suffers in the backward pass, such run time plots provide not much insight into how different frameworks exercise different architectural bottlenecks.}
\label{fig:motivation}
\end{figure}

\begin{figure*}[h]
\centering
\includegraphics[width=\textwidth]{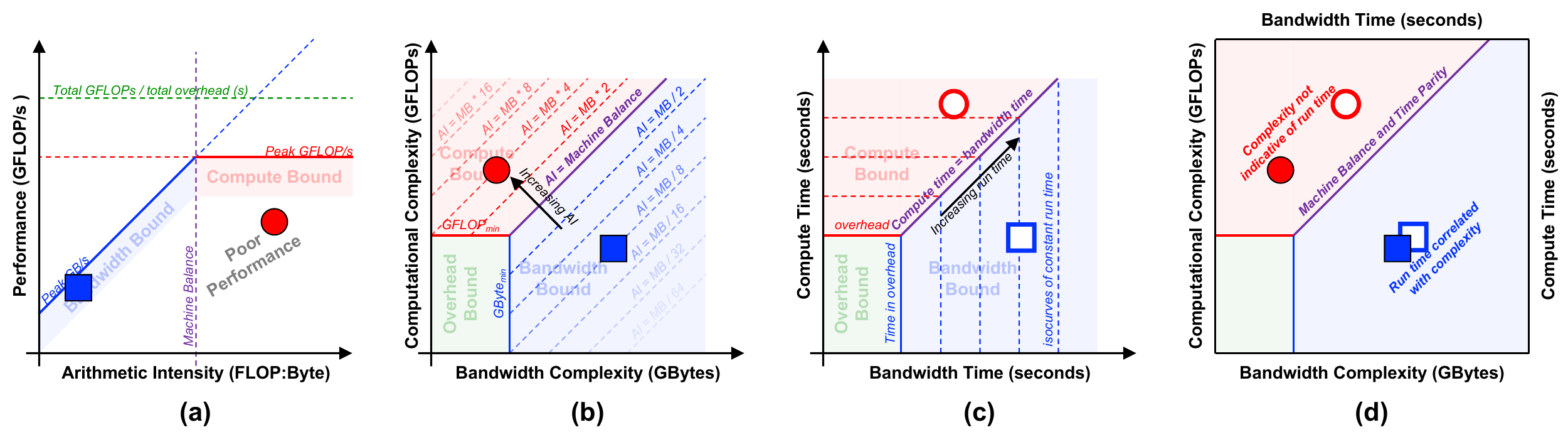}  
\caption{Expanding the Roofline model to incorporate time and complexity. (a) Traditional Roofline, (b) Computational-Bandwidth Complexity, (c) Compute and Bandwidth Time, (d) 4D Complexity-Time Roofline. Note that all plots are in log-log scale.}
\label{fig:intro}
\end{figure*}

As an example, Fig.~\ref{fig:motivation} shows the run time for a 2D convolution kernel implemented in three different deep learning frameworks, PyTorch, TensorFlow v1 and TensorFlow v2, in both forward and backward pass. 
TensorFlow v2 presents a much longer run time in the backward pass due to a different algorithm choice, however, this can not be seen in a run time chart, nor be easily modeled by a traditional Roofline model. 
The traditional Roofline~\cite{williams2009roofline} can provide valuable insights into how different kernels exercise the different architectural components and has been applied to deep learning analysis~\cite{jouppi2017datacenter,javed2019performance}.
But given the algorithm change, even if the computational throughput (FLOP/s) is higher on the Roofline, it does not mean the ultimate run time is better for certain algorithms.

In this paper, we extend the traditional Roofline model to incorporate both computational/bandwidth complexity and run time into the model, and provide a more comprehensive and in-depth methodology for performance analysis of deep learning applications. Throughout the rest of this paper, we will leverage this methodology to analyze Conv2D and LSTM kernels written in PyTorch, TensorFlow~v1, and TensorFlow~v2 as we vary precision and parameters.




\section{Methodology}

\subsection{Roofline Model}
The Roofline performance model is a visually-intuitive performance analysis model used to analyze the performance of applications and computational kernels.
Previous studies have shown that this model can provide valuable insights into performance analysis on various architectures~\cite{doerfler2016applying,koskela2018novel,yang2018toward,yang2020metho,madsen2020timemory,yang2019hierarchical,yang2018empirical,yang2020gpp,ding2019instruction,ibrahim2019performance,powerroofline,alexpowerroofline}.

Roofline is based on two components: 1) characterization of a target architecture in terms of compute (GFLOP/s) and bandwidth (GB/s) potential and 2) characterization of a kernel in terms of ``arithmetic intensity'' (a data locality metric expressed as FLOPs:Bytes) and computational throughput (sustained GFLOP/s).
By construction, the Roofline bound is defined as the minimum of either peak performance or the product of arithmetic intensity and peak bandwidth:
\begin{equation}
\mbox{GFLOP/s} \leq \mbox{min} \left\{ 
  \begin{array}{@{}l@{}}
    \mbox{Peak GFLOP/s}							\\
    \mbox{Peak GB/s} \times \mbox{Arithmetic Intensity}	
  \end{array}
\right.
\label{eq:roofline}
\end{equation}

Fig.~\ref{fig:intro}(a) shows an example of the Roofline model for a nondescript architecture. In this particular case, we incorporate GPU kernel launch overhead into the traditional Roofline model through an additional ceiling (another term in Roofline's min function). This resultant ceiling, defined as total GFLOPs / total overhead, is well above the peak of the machine and thus does not constrain performance.  Had overhead been higher (more kernels) or total GFLOPs lower (smaller kernels), peak performance would have been unattainable. As a result, one can see how the compute (red) and bandwidth (blue) constrain performance and tessellate the resultant 2D locality-performance plane into four regions: bandwidth bound, compute-bound, poor performance, and unattainable performance (above the Roofline).
By scattering two kernels (blue square and red circle) into the resultant space, one can analyze their data locality and performance with the red kernel showing markedly superior data locality and performance (albeit less than machine peak).
\textbf{\textit{Given the superior performance of the red kernel, how could the blue kernel ever have a lower run time?}}

\subsection{Computational and Bandwidth Complexity\label{sec:roof2}}
A common Roofline misconception is that kernels with superior data locality or FLOP rates perform better.  Such kernels may utilize the underlying architecture more efficiently, but Roofline hides whether they require the same number of FLOPs to complete (4$\times$ more work at 2$\times$ the rate).  

Traditionally, \textit{computational complexity} (Big~O notation) has been used to analyze different algorithms by counting the number of operations.  In the case of CNNs, one can choose from several competing algorithms to perform the convolution:  GEMMs, FFTs, and Winograd convolutions.  Each of these methods has a very different computational complexity.  Unfortunately, as Roofline has shown, counting only FLOPs is insufficient in analyzing performance.  To that end, we augment computational complexity with \textit{bandwidth complexity}.  A measure of how much data movement (Bytes) is required to execute an algorithm.

Fig.~\ref{fig:intro}(b) shows that we may create a new model for algorithm complexity.  The axes are defined as the computational and bandwidth complexity of an algorithm.  Isocurves of constant arithmetic intensity are diagonals through the resultant plane with \textit{machine balance} denoting the architecture-specific arithmetic intensity where compute begins to dominate data movement.  We can similarly define a box (complexity less than the product of peak performance and total overhead) wherein GPU kernel launch overhead dominates data movement and compute.  
As in Roofline, we may calculate kernel computational and bandwidth complexity and scatter plot performance in the 2D complexity plane.  Kernels near the origin perform little work while kernels far from the origin have high complexity.  

Generally speaking, as one changes algorithm, parameter, or deep learning framework, one can observe how computational and bandwidth complexity evolve (the trendline/trajectory of an application in the plane).  For example, in Fig.~\ref{fig:intro}(b), we observe both the red and blue kernels are moderately complex with the red kernel having higher computational complexity but lower bandwidth complexity thereby moving it into the compute-bound regime.

Note, the overhead bound region can grow if an algorithm requires more GPU kernel launches or if an alternate architecture has higher overhead.  Similarly, increasing peak performance with the use of tensor cores shifts the machine balance diagonal (transition from bandwidth to compute-bound) to a higher arithmetic intensity.

\subsection{Compute and Bandwidth Time\label{sec:roof3}}
Although orthogonalizing complexity into computational and bandwidth allows users to rapidly and visually analyze the cost of algorithms relative to machine capabilities, it hides perhaps the most important facet: run time.  
The key insight of this paper is that we can remap the computational-bandwidth complexity coordinate system into a 2D time coordinate system: \textit{compute time} and \textit{bandwidth time}. 
In essence, if a kernel is compute-bound in the complexity plane, then we define its compute time as the kernel's run time and its bandwidth time as its run time scaled down by the ratio of arithmetic intensity to machine balance (\ref{eq:computetime}), (\ref{eq:bandwidthtime}).  Conversely, if the kernel is memory-bound, we define its bandwidth time as run time and its compute time as its run time scaled down by the ratio of machine balance to arithmetic intensity.
Implicitly, we assume that the smaller time can be perfectly overlapped with the larger time (i.e. compute time can overlap with bandwidth time).
\begin{equation}
\label{eq:computetime}
\scriptscriptstyle 
\mbox{\scriptsize Compute Time} = min(\frac{\mbox{\scriptsize Computational Complexity}}{\mbox{\scriptsize Peak Performance}}, \frac{\mbox{\scriptsize Bandwidth Complexity}\cdot\mbox{\scriptsize AI}}{\mbox{\scriptsize Peak Performance}})
\end{equation}
\begin{equation}
\label{eq:bandwidthtime}
\scriptscriptstyle 
\mbox{\scriptsize Bandwidth Time} = min(\frac{\mbox{\scriptsize Bandwidth Complexity}}{\mbox{\scriptsize Peak Bandwidth}}, \frac{\mbox{\scriptsize Computational Complexity}}{\mbox{\scriptsize AI}\cdot\mbox{\scriptsize Peak Bandwidth}})
\end{equation}

Fig.~\ref{fig:intro}(c) shows that we can create a compute-bandwidth time model with axes of compute and bandwidth time.  Kernels are scattered into the resultant 2D plane. As before, kernels are overhead-bound if both coordinates are less than the total time spent in GPU kernel launches, bandwidth-bound if bandwidth time exceeds compute time, and compute-bound otherwise.  Total run time can be visualized as isocurves in the 2D time domain.  Thus, if bandwidth-bound, run time increases only if bandwidth time increases, while if compute-bound, run time increases only if compute-time increases.  Kernels nearer the origin have shorter run times while kernels further from the origin have longer run times.

Whereas the traditional Roofline in Fig.~\ref{fig:intro}(a) can only show that the red kernel attained higher performance, the time Roofline in Fig.~\ref{fig:intro}(c) clearly explains why the red kernel is slower --- its run time is determined by its compute time which is greater than the blue kernel's run time (memory time).  This is instantly visualized as it is above the temporal isocurve.

Ultimately, we can combine complexity and time into a single 4-dimensional figure.  Fig.~\ref{fig:intro}(d) shows that rather than trying to plot a 4D tesseract on a 2D piece of paper, for each kernel, we plot the first two coordinates (computational and bandwidth complexity) using a closed symbol on the primary axes and the second two coordinates (compute and memory time) using an open symbol on the secondary axes (each kernel has two symbols).  By scaling compute and bandwidth time by peak GFLOP/s and peak GB/s, we can immediately assess data locality (arithmetic intensity), algorithmic efficiency (complexity), and run time (compute/bandwidth time).  Moreover, the proximity of the open (actual run time) symbol to the closed (complexity) symbol allows us to assess performance (distance from the Roofline).  A kernel whose symbols are widely separated dramatically underperforms its Roofline bound.  A kernel whose symbols are close attains near peak Roofline performance.

\section{Experimental Setup}
\subsection{Hardware and Software Configuration}

Results presented in this paper are obtained from the Cori supercomputer and in particular its GPU partition, at the National Energy Research Scientific Computing Center (NERSC), Lawrence Berkeley National Laboratory (LBNL). 
The GPU partition is primarily deployed for GPU porting, benchmarking, and testing efforts in the NERSC Exascale Science Application Program (NESAP). 
Each node contains two Intel Xeon Gold 6148 Skylake CPUs, 384~GiB DDR4 memory, and 8 NVIDIA V100 GPUs.
Each GPU has 16~GiB of HBM2 memory and 80 SMs, and GPUs on a node are connected in a `hybrid cube-mesh' topology.

On the software side, we have used CUDA 10.2.89, cuDNN 7.6.5, Nsight Compute 2019.5.0, Python 3.7, PyTorch GPU 1.5.0, TensorFlow GPU 1.15.0 (TF1), and TensorFlow GPU 2.2.0 (TF2), for the study in this paper.


\subsection{Roofline Data Collection}
To measure the kernel launch overhead, we have created a micro-benchmark (see~\cite{totocuda2020}) and tested with a wide range of launch parameters, and the average latency we have obtained is 4.2 microseconds on V100s.

For the machine characterization, we have used the Empirical Roofline Toolkit (ERT)~\cite{ert2020} to collect a set of characteristics empirically, such as the HBM bandwidth (828.8~GB/s), single-precision peak (15.16~TFLOP/s), and half-precision peak (29.18~TFLOP/s).
The V100 clock rate, 1.312 GHz, can be collected by using the NVIDIA’s System Management Interface (\texttt{nvidia-smi}) with the command \texttt{nvidia-smi dmon}.
Thus the Tensor Core peak used in this paper is calculated based on the theoretical hardware limits, i.e. $80\times 8\times 1.312\times 4^{3}\times 2 = 107.479 \textrm{ TFLOP/s}$. 
ERT was recently updated with more precision and architecture support, and the implementation details are discussed in~\cite{wang2020pmbs}. Note that the machine balance diagonal presented in this paper is the Tensor Core peak performance divided by the HBM bandwidth (107479 / 828.8 = 129.68).

For application performance data, we utilize a methodology proposed in~\cite{roofline2020} based on \texttt{Nsight Compute} 2020 metrics. 
A set of metrics are collected to measure the kernel run time, computational complexity, and bandwidth complexity (please refer to~\cite{wang2020pmbs,yang2020metho} for more details). Throughout this paper, computational complexities are treated equally thus precision-agnostic.



\subsection{2D Convolution}
In CNN models, the 2D Convolution (or Conv2D) layer are usually the most compute-intensive components in deep learning applications~\cite{gu2018recent}. In a forward pass, it performs a convolution of an input tensor and a convolution kernel (or filter), and generates a new tensor as the output. For an input tensor $A$ of shape $N \times H \times W \times C$ (where $N$ is the number of samples in the batch, $H$ and $W$ are the height and width, and $C$ is the number of channels), and a convolution kernel $K$ of shape $K_H$, $K_W$, $C$, $C'$, the output tensor $B$ is of shape $N \times H' \times W' \times C'$ with $H'=H-K_{H}+1$, $W'= W-K_{W}+1$, and the convolution is defined as 
\begin{equation}
\label{eq:conv2d}
    B_{nhwc} = \sum_{m=0}^{C-1} \sum_{K_{h}=0}^{K-{H}-1} \sum_{K_{w}=0}^{K_{W}-1} A_{nh+K_{h} w+w_{h}m} K_{k_{h}k_{w}mc}
\end{equation}

After the convolution, there is usually an offset value added to each element, called the ``bias", and a non-linear activation function such as ReLU~\cite{hahnloser2000digital} is applied, but this is usually light on computation.
In the backward pass, the gradients are computed with respect to all trainable variables, and it is usually more computationally intensive than the forward pass.



In this study, we will focus on the performance characteristics of a typical Conv2D kernel in one of the most widely used networks for image analysis, ResNet50~\cite{he2016deep}. Both single-precision and half-precision data will be evaluated as they are the most relevant data types in deep learning. To make a fair comparison among different frameworks, we initialize the convolution kernel beforehand and focus only on the computation loop itself. As a result, filling input data with random number generators or kernel initialization is not included in the measurement. To achieve this, the \texttt{profile-from-start} option is disabled in \texttt{Nsight Compute} and CuPy~\cite{okuta2017cupy} is used to explicitly restrict the profiling region to include the computation loop only. Also, we set up a warm-up loop with 5 iterations before the computation loop to get rid of unoptimized auto-tuning kernels in TensorFlow. The profiling results include on average over 20 iterations of the computation kernel.

\subsection{LSTM (Long Short-Term Memory neural networks)}
\label{sec:lstm}
LSTMs are in the family of recurrent neural networks (RNNs) and usually used for speech and language processing as well as time-series sequential data. For sequential data in general, recurrent models have the capabilities to maintain hidden state $h_{t-1}$ representing features in the previous time steps. Hence, the prediction at the current step $o_t$ should take account of both the current input $x_t$ and the hidden state feature from the previous time step $h_{t-1}$. 
\begin{equation}
\label{eq:lstm1}
\begin{aligned}
    f_t & = \sigma ( W_{f} [ h_{t-1} , x_t ] + b_f )\\
    i_t & = \sigma ( W_i [h_{t-1} , x_t ] + b_i )\\
    \hat{C_t} & = tanh ( W_C [h_{t-1}, x_t] + b_C )\\
    C_t & = f_t * C_{t-1} + i_t * \hat{C_t}\\
    o_t & = \sigma ( W_o [ h_{t-1} , x_t ] + b_o )\\
    h_t & = o_t * tanh(C_t)
\end{aligned}
\end{equation}

The main advantage of LSTMs over basic RNNs is their capability to learn long-term dependencies during training, which helps to overcome the vanishing gradient problem. Besides the hidden state feature $h_t$ being passed through the network in basic RNNs, LSTMs introduce an additional data flow $C_t$ carrying the information which combines the current input $x_t$ and hidden feature from past steps $h_{t-1}$. The gates related to this data flow are used to control the information being sent to the next time-step. Ideally, the carry data flow modulates the next output $o_t$ and the next state $h_t$, and it allows for

Compared to Conv2D kernels which can be fully parallelized, operations in an LSTM cell have dependencies and part of them will only be executed sequentially. For example, the computation of gates, i.e., $f_t$, $i_t$, $\hat{C_t}$ and $o_t$, takes the same input and can be performed in parallel without issues. However, the computation of current cell state $C_t$ depends on gate values (\ref{eq:lstm1}) thus has to be performed in serial. In the end, the hidden state $h_t$ depends on the cell state so theoretically there are at least three steps run sequentially to calculate one LSTM cell. Except for computing $C_t$ which is similar to a common convolution operation, other operations involved in this process are clearly less compute-intensive than in a Conv2D kernel. Moreover, since LSTM cells at different time steps have to be calculated sequentially as well, we can conclude that it has less parallelism than Conv2D. Also, we can expect that LSTM kernels have a low FLOP/s performance and high data movement compared to Conv2D.
\begin{figure*}[!htp]
\centering
\begin{subfigure}{.32\linewidth}
  \centering
  \includegraphics[width=\linewidth]{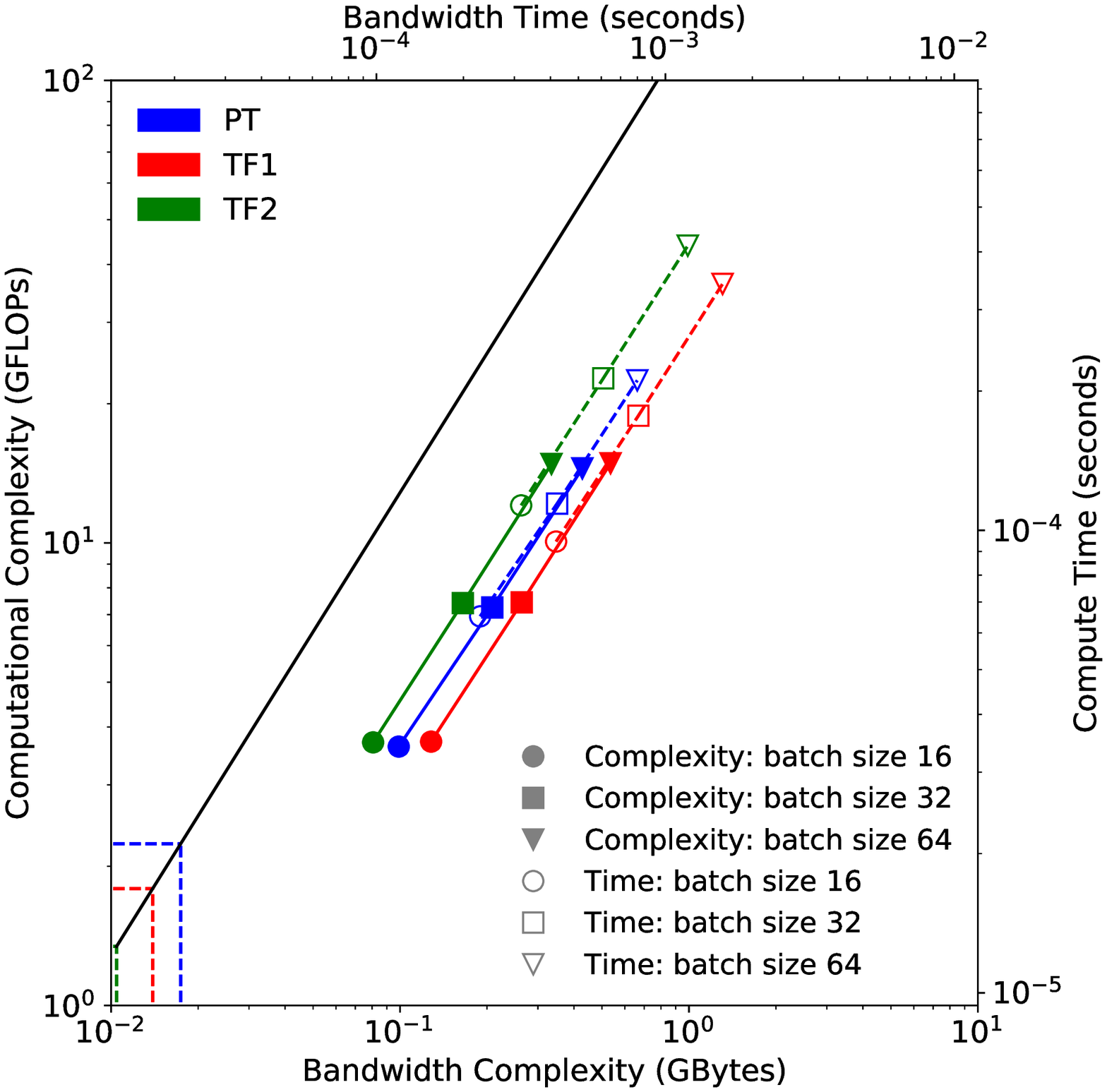}  
  \caption{FP16 forward pass}
\end{subfigure}
\begin{subfigure}{.32\linewidth}
  \centering
  \includegraphics[width=\linewidth]{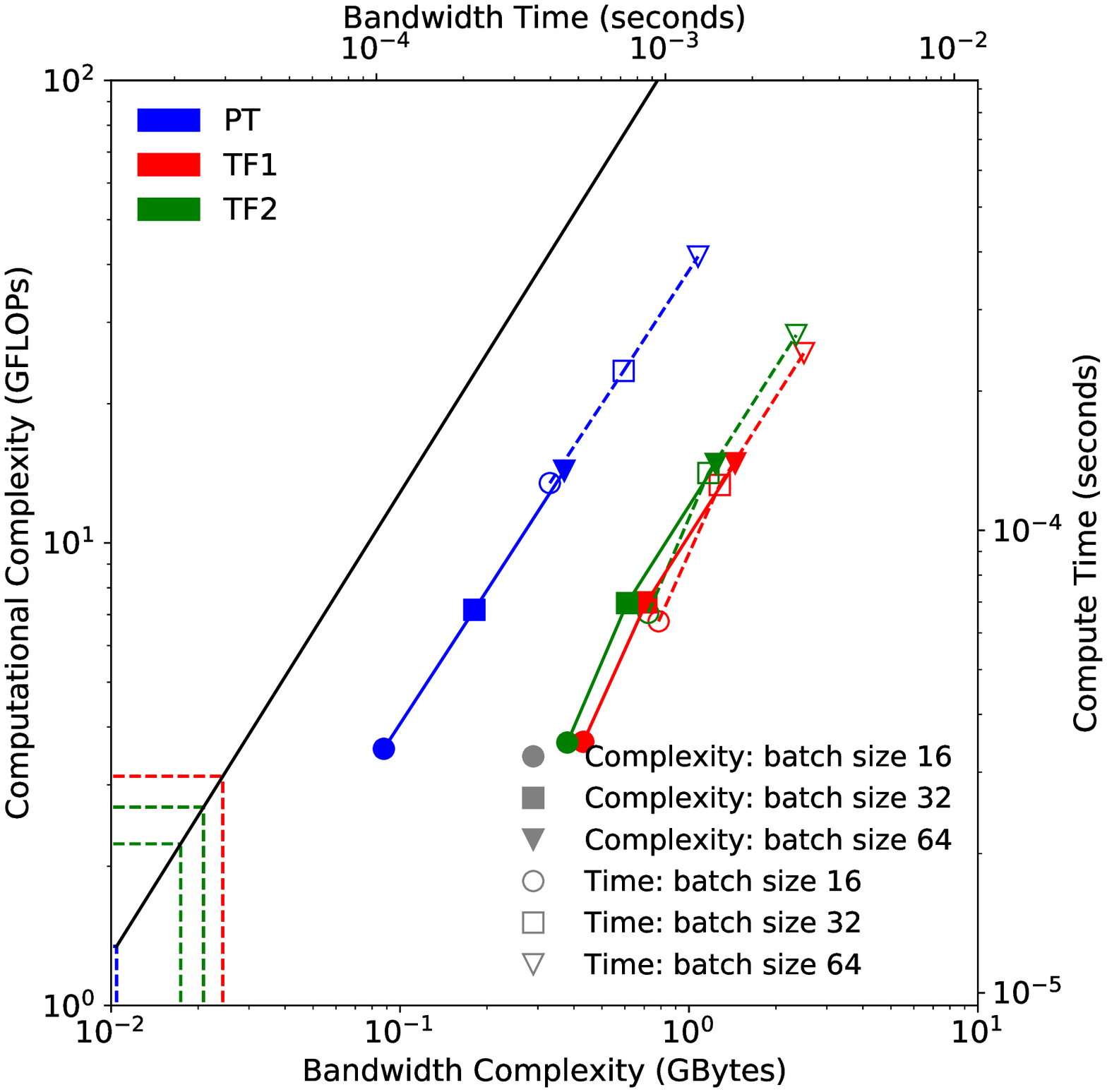} 
  \caption{FP32 forward pass}
\end{subfigure}
\begin{subfigure}{.32\linewidth}
  \centering
  \includegraphics[width=\linewidth]{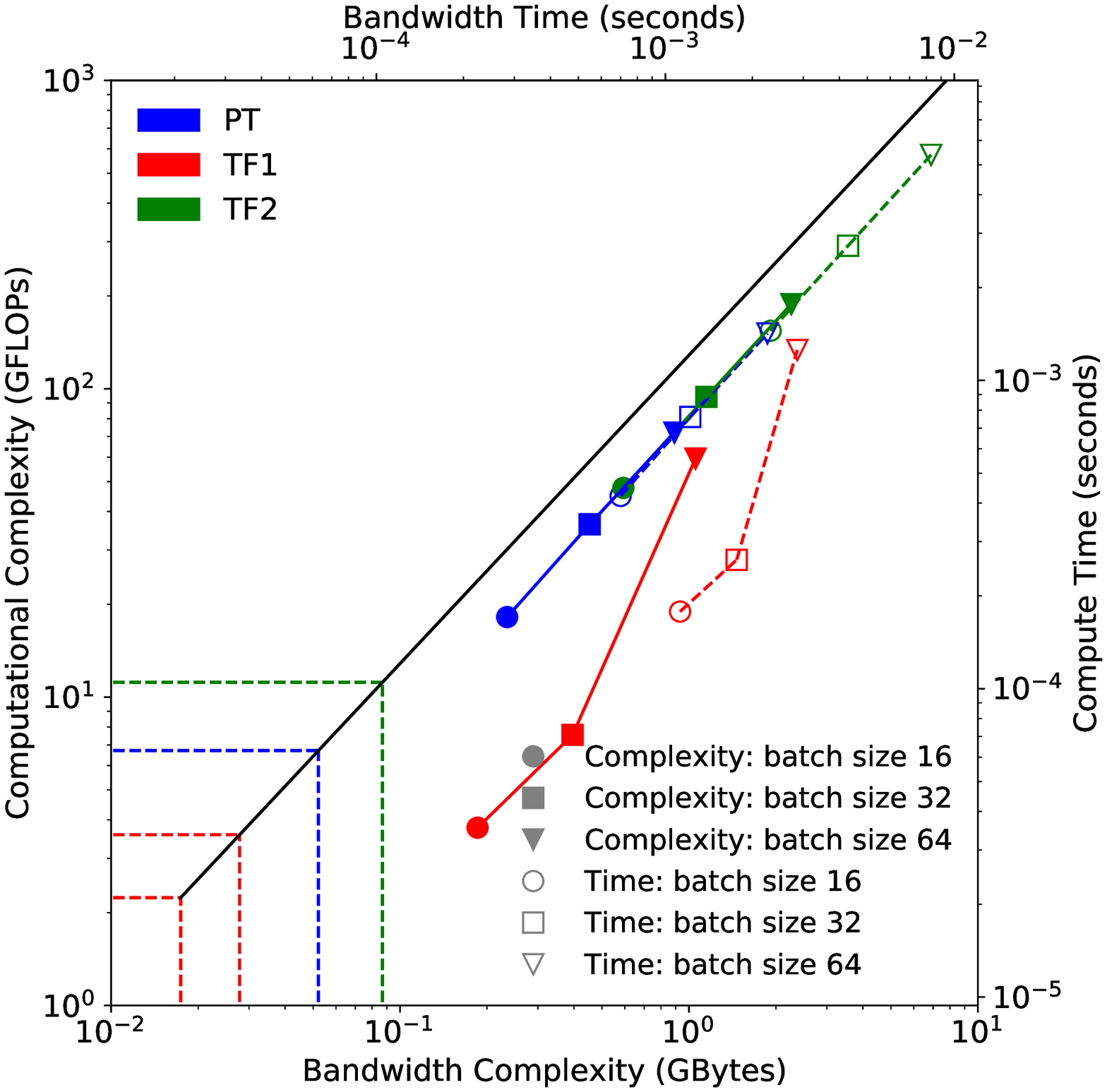}  
  \caption{FP16 backward pass}
\end{subfigure}
\caption{Performance effects by varying the batch size in a Conv2D kernel. The two TensorFlow frameworks tend to use the same underlying algorithms and deliver very similar run time performance. PyTorch presents the same bandwidth complexity in half and single precisions, suggesting the use of an implicit type conversion mechanism.}
\label{fig:conv2d_batch}
\end{figure*}
Profiling set up for LSTM is the same as Conv2D where we evaluate single-precision and half-precision data respectively and focus on the pure computation loop without kernel or data initialization included.

\section{Results}
In this section, we will examine the effects of different input parameters on the performance of Conv2D and LSTM kernels using our time-based Roofline methodology. 
We have implemented these kernels in three different frameworks, PyTorch, TF1, and TF2, and trendlines or trajectory lines will be presented for results from the same framework and precision format (FP32 or FP16). 
Solid lines connect solid symbols to represent computational and bandwidth complexity and dashed lines connect open symbols to show run time (see description in Sec.~\ref{sec:roof2} and~\ref{sec:roof3}). Note that for each algorithm, the overall overhead is calculated as the average overhead (4.2 ms) multiplies with the total number of invocations involved in the algorithm. 
While a range of model parameters will be examined, only one will vary at a time. Due to a large number of results, only a select set is presented in this section. For more complete results, please see~\cite{kernels2020}.

\subsection{2D Convolution}
The four parameters we study for the Conv2D kernel are the batch size, number of filters, kernel size, and stride size. Unless otherwise stated, the default batch size is 16, stride size is 2, input tensor shape is $112\times 112\times 64$, and the kernel size is $3\times 3\times 64\times C'$, where $C'$ is the number of filters defined in (\ref{eq:conv2d}), $C'=64$ by default.

\subsubsection{Batch Size}

Fig.~\ref{fig:conv2d_batch} shows the time-based Roofline charts when varying the batch size from 16, 32 to 64. In the forward pass, Fig.~\ref{fig:conv2d_batch}(a) and~\ref{fig:conv2d_batch}(b) specifically, the arithmetic intensity (AI) is preserved for each framework, PyTorch, TF1, or TF2, and that suggests that the same underlying algorithm is used regardless of the batch size.
The TF1 and TF2 results are highly correlated since their trendlines are always identical, with slight shifting.
This shifting is caused by an extra \texttt{FillPhiloxRandomKernelLaunch} call invoked in TF1 but not in TF2, and as a result, TF2 performs slightly better than TF1 in both run time and compute throughput. 
PyTorch, however, presents a different pattern.
In theory, when switching from FP16 to FP32, a 2$\times$ increase in data movement is expected - TF1 and TF2 show a 3$\times$ increase, but the bandwidth complexity for PyTorch stays almost constant, indicating that there is possibly an auto type-conversion mechanism taking place to convert the data to lower precisions implicitly in order to improve the overall efficiency.

In the backward pass, the kernel performance is not as uniform as in the forward pass. One obvious difference between different frameworks is the number of kernel invocations and this can be seen by the sizes of the overhead boxes in Fig.~\ref{fig:conv2d_batch}(c).
TF2 (green) has significantly higher net launch overhead than TF1 (red), while PyTorch (blue) is in the middle. 
The difference between TF1 and TF2 could be caused by the extensive use of Eigen~\cite{guennebaud2010eigen} functions in TF2 which has contributed to the overall launch overhead.
In terms of kernel run time, PyTorch outperforms TF1 and TF2 as it has fewer zero-AI kernels used for data padding, shuffling, and rearranging, than in TF1 and TF2.
Out of the three frameworks, TF1 is the only one that does not use Tensor Core operations at smaller batch sizes, resulting in lower arithmetic intensity; however, its auto-tuning mechanism does direct to Tensor Core implementations as the batch size increases.


\subsubsection{Number of Output Channels}
\begin{figure}[!htp]
\centering
\includegraphics[width=0.9\linewidth]{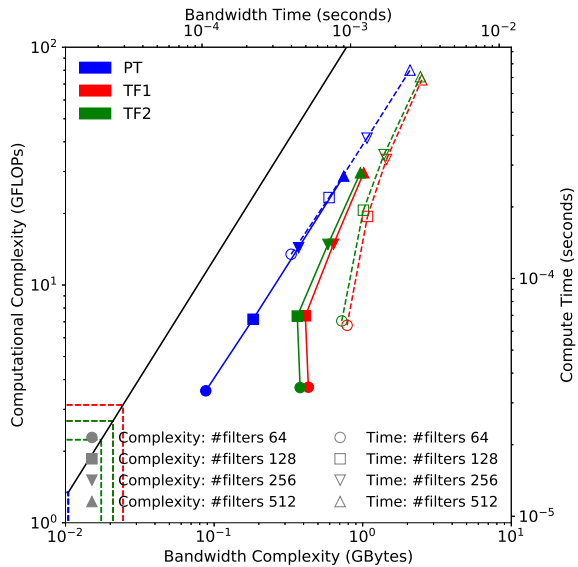}
\caption{The forward pass of an FP32 Conv2D kernel as the number of filters varies. The arithmetic intensity of the PyTorch kernel remains constant.}
\label{fig:conv2d_filters_forward}
\end{figure}
In the forward pass, TF1 and TF2 deliver a very similar performance since both of their trendlines are highly coherent (Fig.~\ref{fig:conv2d_filters_forward}). Besides, compared to relatively constant FLOP/s performance and AIs showed in Fig.~\ref{fig:conv2d_batch}(b), we can find that a larger number of output channels will result in a higher FLOP rate and higher arithmetic intensity for both TF1 and TF2. It's simply because the kernel is much smaller than the input tensor so compared to the increase of memory space to store more channels, the increase of FLOP counts brought by extra channels is tremendously more significant. However, the same conclusion cannot be applied to describe single-precision PyTorch behavior since the performance rarely changed. The bandwidth complexity is proportional to its computational complexity. In half-precision, all three frameworks have very similar trendlines and TF2 outperforms the other two due to less data movement. We find that three frameworks use the same 2D convolution kernel but PyTorch invokes an extra \texttt{nchwToNhwcKernel} function to perform data transpose.
\begin{figure}[t]
\centering
\includegraphics[width=0.9\linewidth]{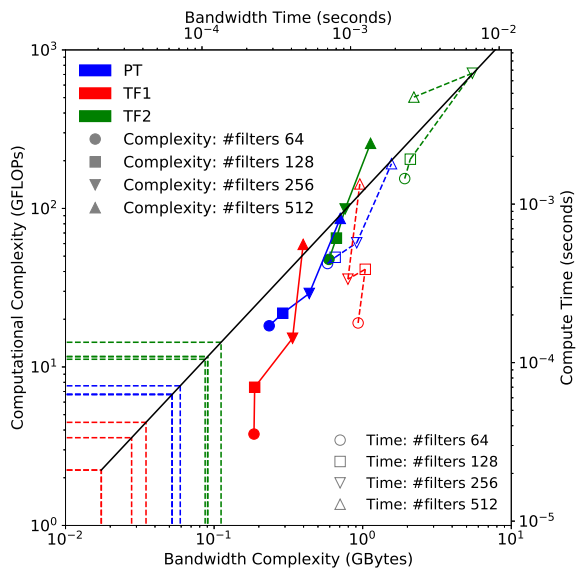}
\caption{The backward pass of an FP16 Conv2D kernel as the number of filters varies. Algorithmic choices are in constant change.}
\label{fig:conv2d_filters_backward}
\end{figure}

\begin{figure}[t]
\includegraphics[width=0.9\linewidth]{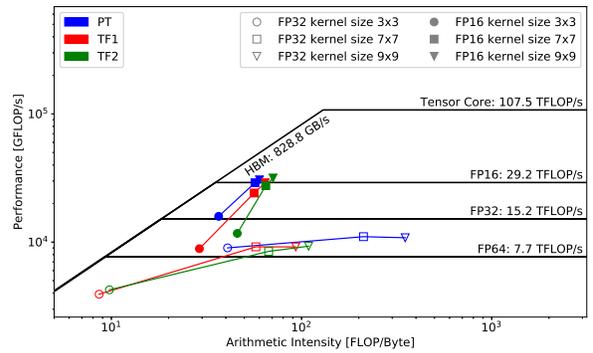}  
\caption{The classic Roofline plot highlights FLOP/s performance that cannot be reflected by the time-based Roofline. PyTorch single-precision FLOP/s performance remains constant implying it's saturating certain hardware resources.}
\label{fig:roofline_kernelsize}
\end{figure}
The backward pass performance is highly diverse based on the number of filters, the precision, and the framework (see Fig.~\ref{fig:conv2d_filters_backward}). In general, TF1 has the lowest computational complexity among the three frameworks and is the least concerned by the kernel launch latency. In half-precision, we can conclude that the choice of algorithms across all three frameworks seems very sensitive to the number of output channels since they never reuse the same set of kernels for a different parameter in this case. In the end, with 512 filters, kernels in all three frameworks are transferred from memory-bound to compute-bound. TF2 delivers the highest FLOP/s performance and arithmetic intensity and also requires the longest run time.


\subsubsection{Kernel Size}
Similar to varying the number of output filters, the increase of kernel size should also augment computation complexity for the same amount of data transfer so it will result in a higher arithmetic intensity and higher FLOP/s performance. Also, constant data movement across different kernel sizes can be expected since the kernel is usually much smaller than the input tensor.
Profiling results approve the above statement except that PyTorch delivers a constant FLOP/s performance in single-precision (see Fig.~\ref{fig:roofline_kernelsize}). This indicates that the algorithm has saturated certain hardware resources thus no matter how kernel size increases, the performance will not increase anymore. To improve this situation, the possibility of using an alternative convolution algorithm should be investigated.

\subsubsection{Stride Size}
\begin{figure}[t]
\centering
  \centering
  \includegraphics[width=0.9\linewidth]{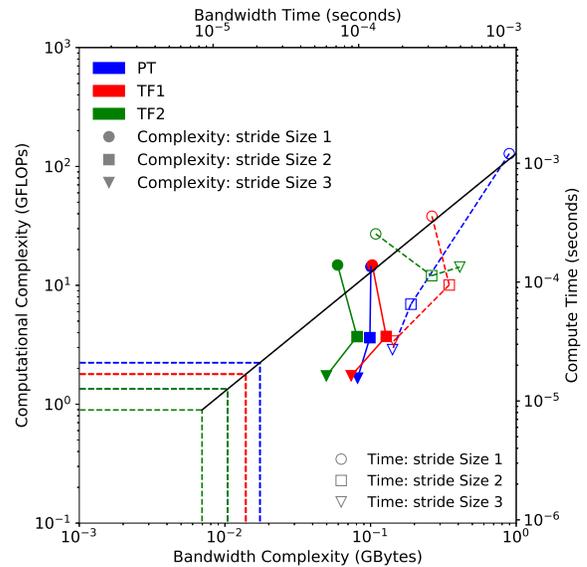}
\caption{Performance effects by varying the stride size in a FP16 Conv2D forward pass.}
\label{fig:conv2d_stridesize}
\end{figure}
Increasing stride size signifies a lower computational complexity for the same 
bandwidth complexity since more input data is masked out during the computation.
Profiling results (see Fig.~\ref{fig:conv2d_stridesize}) are coherent with expectation since bandwidth complexities are relatively constant compared to computational complexities. Besides, it should be noted that lower computational complexity doesn't necessarily mean fewer invocations so the kernel becomes more overhead-bound when the stride size increases.
\begin{figure}[!htp]
\centering
\includegraphics[width=0.9\columnwidth]{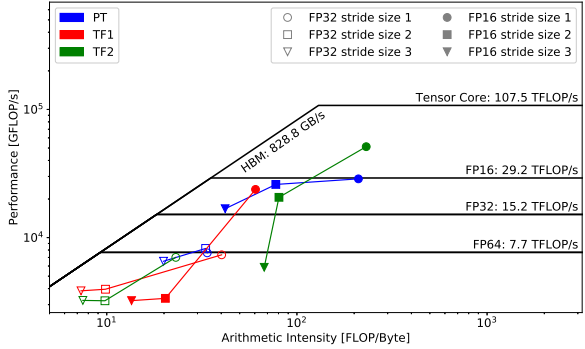}
\caption{Sudden performance drop is often related to poor algorithmic choices. TF2 delivers a significantly lower performance when \texttt{stride\_size=3} because it uses single-precision algorithm for half-precision data.}
\label{fig:roofline_stridesize}
\end{figure}

In half-precision with stride size equaling to 1, PyTorch and TF2 codes are compute-bound while TF1 is memory-bound. Furthermore, PyTorch and TF2 both have a FLOP/s performance higher than TF1 (Fig.~\ref{fig:roofline_stridesize}). Also, we can find that TF1 performance drops largely with larger stride sizes. The reason is that TF1 tends to use a non-Tensor Core algorithm, \texttt{wgrad\_alg0\_engine}, for stride size equals to 2 and 3, which delivers a peak performance of only 5 TFLOPs per second. The same as the forward pass, TF2 drops performance with stride size equals to 3 because of the use of single-precision algorithms for half-precision data.

\subsection{LSTM}
As eluded to in Section~\ref{sec:lstm}, LSTM is computationally different from convolutional kernels since it is not dominated by compute-intensive GEMM operations. In this section, we will vary four parameters in the LSTM kernel and discuss how they impact performance. The four parameters are batch size, sequence length, number of features, and hidden feature size, with default values as 16, 16, 32, and 16 respectively.

\subsubsection{Batch Size}
\begin{figure}[!htp]
\centering
  \includegraphics[width=0.9\linewidth]{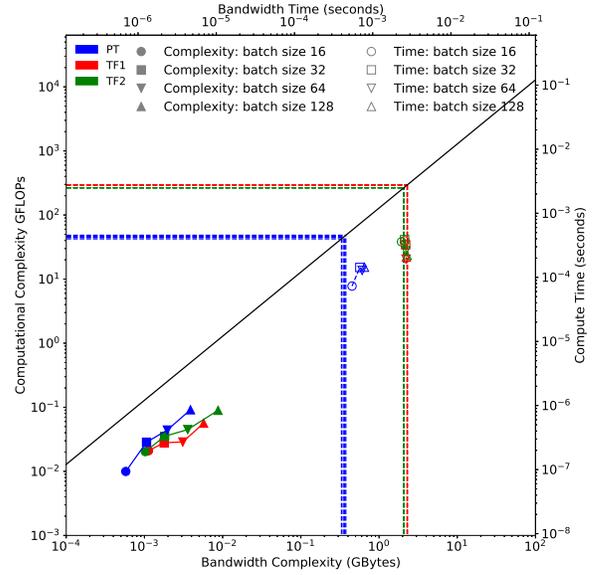}  
\caption{Performance effects of batch size in half-precision LSTM backward pass. The kernel is always overhead-bounded and run time remains the same no matter how we vary the batch size.}
\label{fig:lstm_batchsize}
\end{figure}
Fig.~\ref{fig:lstm_batchsize} shows the time-based Roofline of LSTM when running with four different batch sizes, 16, 32, 64 and 128. The result is very different from convolution kernels and the computation is always overhead-bound regardless of the batch size and forward/backward pass. It is clear that a larger batch size will bring more FLOPs and data movement to perform thus brings a visible increase in computational complexity. However, since the kernel is overhead-bound in nature, trivial improvements on AIs or FLOP/s performance will not affect kernel run time. The run time is only a function of launch latency and the number of kernels invoked in the computation.
In theory, the kernel should perform at least $1.06e{14} \times 4.2e{-6} = 0.4452$ GFLOPs which is about 3000 samples in one batch ($batch\_size = 3000$) in the current forward pass test to be not bounded by the kernel launch latency.

We noticed that TF1 and TF2 invoke significantly more kernels than PyTorch since their overhead boxes are larger than that of PyTorch (Fig.~\ref{fig:lstm_batchsize}). Detailed profiling shows that in the forward pass with default parameters, there are 36 invocations in the PyTorch forward pass where 16 of them are single precision GEMM operations, \texttt{gemmSN\_TN\_kernel}. We also find that this GEMM kernel is always invoked along with \texttt{LSTM\_elementWise\_fp} thus this function pair represents 32 invocations in total. In theory, 16 convolutions are required to compute 16 time steps but PyTorch invokes another two half-precision S884 GEMM kernels on top of the 16 single-precision convolutions. In the end, there are only two zero-AI kernels for pure data movement purposes. For the same test case, TF1 and TF2 have 277 and 243 invocations respectively. Note that Eigen functions are observed to be widely used in these two frameworks and the drawback is that Eigen tends to invoke a large number of memory-intensive \texttt{EigenMetaKernel} with very low FLOP/s performance (0 $\sim$ 1 GFLOPs/s). As for the convolution kernel in TF1 and TF2, we can find that TF1 is always using the Tensor Core S884 GEMM algorithm and TF2 employs \texttt{gemmSN\_NN\_kernel} which is similar to PyTorch but with 2$\times$ more invocations. In terms of FLOP/s performance, TF2 has similar trendlines to PyTorch other than TF1. In conclusion, PyTorch could be further optimized by applying a more proper half-precision GEMM kernel. Though TF1 has a better choice of the GEMM algorithm, it loses its efficiency by invoking too many housekeeping kernels. TF2 doesn't use the proper half-precision GEMM algorithms and it performs redundant GEMM operations as well. However, it invokes less memory-intensive kernels than TF1 thus TF1 and TF2 finally attain the same run time.

Similar conclusions in this section can be applied to changing the number of features and hidden size as well. In all three cases, kernel invocations are pretty constant within the same framework no matter how the parameter varies and PyTorch is always the most efficient framework due to the least kernel invocations.

\subsubsection{Sequence Length}
\begin{figure}[t]
\centering
  \includegraphics[width=0.9\linewidth]{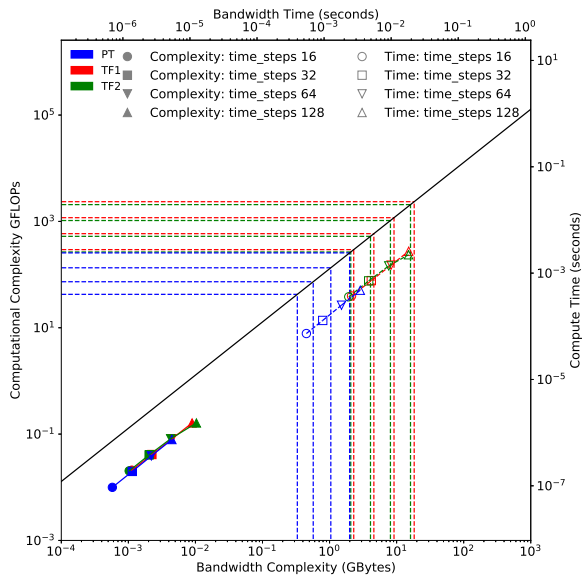}  
\caption{Performance effects of sequence length in LSTM. Long sequence length implies repeating the same computation for more times. Kernel AI remains constant and run time is proportional to sequence length.}
\label{fig:lstm_time_timesteps}
\end{figure}

Among four parameters we tested for an LSTM kernel, changing sequence length is the only case where run time and kernel invocations will vary along with the parameter value. Equation~(\ref{eq:lstm1}) shows the computation to perform for each time step and increasing the sequence length indicates how many times this process will be repeated in a sequence. The sequence cannot be parallelized thus kernel invocations should be proportional to the sequence length. Fig.~\ref{fig:lstm_time_timesteps} confirms this suspicion. It shows the time-based Roofline with four sequence lengths: 16, 32, 64, and 128. Besides, it can be found that all solid symbols and their corresponding open symbols are lying on the same AI diagonals meaning the algorithmic choice is constant. As a result, we will find the FLOP/s performance will remain the same when varying the sequence length. PyTorch outperforms the other two as it moves less data, performs fewer FLOPs, and requires fewer kernel invocations.



\section{Conclusion}

In this paper, we proposed a new time-based Roofline model as an extension to the traditional Roofline, to model both computational/bandwidth complexity and run time. This new model helps users understand important performance characteristics such as algorithm changes and high kernel launch overhead that are pertinent to deep learning applications and could not be modeled by the traditional Roofline.

We have applied this new methodology to two representative deep learning kernels, 2D convolution (from CNN) and LSTM (from RNN), to demonstrate the use of this model and its effectiveness.  
Three common frameworks, PyTorch, TensorFlow v1, and TensorFlow v2, are compared. 
Results show that the use of the Tensor Core pipeline and cuDNN functions are critical to achieving optimal performance on NVIDIA V100 GPUs. 2D convolutions are compute-intensive and benefit well from the Tensor Core utilization. LSTM is computationally distinct from 2D convolution and kernel performance is highly bounded by its launch overhead. In both test cases, PyTorch outperforms the other two frameworks in terms of run time as it moves less data, performs fewer FLOPs and requires fewer kernel invocations.

Ultimately, we show that the new time-based Roofline model can provide a more systematic way of understanding deep learning application performance and comparing different implementations that may have similar FLOP/s performance and arithmetic intensity but are completely different in computational and bandwidth complexity. By applying it into the design and development of new deep learning architectures or frameworks, architects and engineers can rely on one single model to identify the performance issue without going through a large collection of different performance metrics. Finally yet importantly, it is a generic methodology for performance analysis thus can be perfectly applied on non deep learning use cases as well.
\section*{Acknowledgements}
This material is based upon work supported by the Advanced Scientific Computing Research Program in the U.S. Department of Energy, Office of Science, under Award Number DE-AC02-05CH11231.
This research used resources of the National Energy Research Scientific Computing Center (NERSC) which is supported by the Office of Science of the U.S. Department of Energy under Contract No. DE-AC02-05CH11231.
We thank NVIDIA Corporation for their willingness to answer our myriad of questions on Nsight metrics.

\bibliographystyle{IEEEtran}
\bibliography{IEEEabrv,bibfile}

\begin{thebibliography}{10}
\providecommand{\url}[1]{#1}
\csname url@samestyle\endcsname
\providecommand{\newblock}{\relax}
\providecommand{\bibinfo}[2]{#2}
\providecommand{\BIBentrySTDinterwordspacing}{\spaceskip=0pt\relax}
\providecommand{\BIBentryALTinterwordstretchfactor}{4}
\providecommand{\BIBentryALTinterwordspacing}{\spaceskip=\fontdimen2\font plus
\BIBentryALTinterwordstretchfactor\fontdimen3\font minus
  \fontdimen4\font\relax}
\providecommand{\BIBforeignlanguage}[2]{{%
\expandafter\ifx\csname l@#1\endcsname\relax
\typeout{** WARNING: IEEEtran.bst: No hyphenation pattern has been}%
\typeout{** loaded for the language `#1'. Using the pattern for}%
\typeout{** the default language instead.}%
\else
\language=\csname l@#1\endcsname
\fi
#2}}
\providecommand{\BIBdecl}{\relax}
\BIBdecl

\bibitem{deng2009imagenet}
J.~Deng, W.~Dong, R.~Socher, L.-J. Li, K.~Li, and L.~Fei-Fei, ``Imagenet: A
  large-scale hierarchical image database,'' in \emph{2009 IEEE conference on
  computer vision and pattern recognition}.\hskip 1em plus 0.5em minus
  0.4em\relax Ieee, 2009, pp. 248--255.

\bibitem{lecun1995convolutional}
Y.~LeCun, Y.~Bengio \emph{et~al.}, ``Convolutional networks for images, speech,
  and time series,'' \emph{The handbook of brain theory and neural networks},
  vol. 3361, no.~10, p. 1995, 1995.

\bibitem{goodfellow2014generative}
I.~Goodfellow, J.~Pouget-Abadie, M.~Mirza, B.~Xu, D.~Warde-Farley, S.~Ozair,
  A.~Courville, and Y.~Bengio, ``Generative adversarial nets,'' in
  \emph{Advances in neural information processing systems}, 2014, pp.
  2672--2680.

\bibitem{graves2013hybrid}
A.~Graves, N.~Jaitly, and A.-r. Mohamed, ``Hybrid speech recognition with deep
  bidirectional lstm,'' in \emph{2013 IEEE workshop on automatic speech
  recognition and understanding}.\hskip 1em plus 0.5em minus 0.4em\relax IEEE,
  2013, pp. 273--278.

\bibitem{cho2014learning}
K.~Cho, B.~Van~Merri{\"e}nboer, C.~Gulcehre, D.~Bahdanau, F.~Bougares,
  H.~Schwenk, and Y.~Bengio, ``Learning phrase representations using rnn
  encoder-decoder for statistical machine translation,'' \emph{arXiv preprint
  arXiv:1406.1078}, 2014.

\bibitem{manning2014stanford}
C.~D. Manning, M.~Surdeanu, J.~Bauer, J.~R. Finkel, S.~Bethard, and
  D.~McClosky, ``The stanford corenlp natural language processing toolkit,'' in
  \emph{Proceedings of 52nd annual meeting of the association for computational
  linguistics: system demonstrations}, 2014, pp. 55--60.

\bibitem{he2016deep}
K.~He, X.~Zhang, S.~Ren, and J.~Sun, ``Deep residual learning for image
  recognition,'' in \emph{Proceedings of the IEEE conference on computer vision
  and pattern recognition}, 2016, pp. 770--778.

\bibitem{torchprofiler}
\BIBentryALTinterwordspacing
{PyTorch Profiler}. [Online]. Available:
  \url{https://pytorch.org/tutorials/recipes/recipes/profiler.html}
\BIBentrySTDinterwordspacing

\bibitem{tfprofiler}
\BIBentryALTinterwordspacing
{TensorFlow Profiler}. [Online]. Available:
  \url{https://www.tensorflow.org/tensorboard/tensorboard\_profiling\_keras}
\BIBentrySTDinterwordspacing

\bibitem{dlprof}
\BIBentryALTinterwordspacing
{NVIDIA Deep Learning Profiler (DLProf)}. [Online]. Available:
  \url{https://docs.nvidia.com/deeplearning/frameworks/dlprof-user-guide/}
\BIBentrySTDinterwordspacing

\bibitem{jorda2019performance}
M.~Jord{\`a}, P.~Valero-Lara, and A.~J. Pe{\~n}a, ``Performance evaluation of
  cudnn convolution algorithms on nvidia volta gpus,'' \emph{IEEE Access},
  vol.~7, pp. 70\,461--70\,473, 2019.

\bibitem{ren2019performance}
Y.~Ren, S.~Yoo, and A.~Hoisie, ``Performance analysis of deep learning
  workloads on leading-edge systems,'' in \emph{2019 IEEE/ACM Performance
  Modeling, Benchmarking and Simulation of High Performance Computer Systems
  (PMBS)}.\hskip 1em plus 0.5em minus 0.4em\relax IEEE, 2019, pp. 103--113.

\bibitem{Wang_2020_CVPR}
Y.~Wang, X.~Wei, F.~Liu, J.~Chen, Y.~Zhou, W.~Shen, E.~K. Fishman, and A.~L.
  Yuille, ``Deep distance transform for tubular structure segmentation in ct
  scans,'' in \emph{IEEE/CVF Conference on Computer Vision and Pattern
  Recognition (CVPR)}, June 2020.

\bibitem{Dolhasz_2020_CVPR}
A.~Dolhasz, C.~Harvey, and I.~Williams, ``Learning to observe: Approximating
  human perceptual thresholds for detection of suprathreshold image
  transformations,'' in \emph{IEEE/CVF Conference on Computer Vision and
  Pattern Recognition (CVPR)}, June 2020.

\bibitem{Schops_2020_CVPR}
T.~Schops, V.~Larsson, M.~Pollefeys, and T.~Sattler, ``Why having 10,000
  parameters in your camera model is better than twelve,'' in \emph{IEEE/CVF
  Conference on Computer Vision and Pattern Recognition (CVPR)}, June 2020.

\bibitem{shao2017performance}
L.~Shao, Z.~Cai, L.~Liu, and K.~Lu, ``Performance evaluation of deep feature
  learning for rgb-d image/video classification,'' \emph{Information Sciences},
  vol. 385, pp. 266--283, 2017.

\bibitem{carneiro2018performance}
T.~Carneiro, R.~V.~M. Da~N{\'o}brega, T.~Nepomuceno, G.-B. Bian, V.~H.~C.
  De~Albuquerque, and P.~P. Reboucas~Filho, ``Performance analysis of google
  colaboratory as a tool for accelerating deep learning applications,''
  \emph{IEEE Access}, vol.~6, pp. 61\,677--61\,685, 2018.

\bibitem{williams2009roofline}
S.~Williams, A.~Waterman, and D.~Patterson, ``Roofline: An insightful visual
  performance model for floating-point programs and multicore architectures,''
  Lawrence Berkeley National Lab.(LBNL), Berkeley, CA (United States), Tech.
  Rep., 2009.

\bibitem{jouppi2017datacenter}
N.~P. Jouppi, C.~Young, N.~Patil, D.~Patterson, G.~Agrawal, R.~Bajwa, S.~Bates,
  S.~Bhatia, N.~Boden, A.~Borchers \emph{et~al.}, ``In-datacenter performance
  analysis of a tensor processing unit,'' in \emph{Proceedings of the 44th
  Annual International Symposium on Computer Architecture}, 2017, pp. 1--12.

\bibitem{javed2019performance}
M.~H. Javed, K.~Z. Ibrahim, and X.~Lu, ``Performance analysis of deep learning
  workloads using roofline trajectories,'' \emph{CCF Transactions on High
  Performance Computing}, vol.~1, no.~3, pp. 224--239, 2019.

\bibitem{doerfler2016applying}
D.~Doerfler, J.~Deslippe, S.~Williams, L.~Oliker, B.~Cook, T.~Kurth, M.~Lobet,
  T.~Malas, J.-L. Vay, and H.~Vincenti, ``Applying the roofline performance
  model to the intel xeon phi knights landing processor,'' in
  \emph{International Conference on High Performance Computing}.\hskip 1em plus
  0.5em minus 0.4em\relax Springer, 2016, pp. 339--353.

\bibitem{koskela2018novel}
T.~Koskela, Z.~Matveev, C.~Yang, A.~Adedoyin, R.~Belenov, P.~Thierry, Z.~Zhao,
  R.~Gayatri, H.~Shan, L.~Oliker, J.~Deslippe, R.~Green, and S.~Williams, ``{A
  Novel Multi-Level Integrated Roofline Model Approach for Performance
  Characterization},'' in \emph{International Conference on High Performance
  Computing}.\hskip 1em plus 0.5em minus 0.4em\relax Springer, 2018, pp.
  226--245.

\bibitem{yang2018toward}
C.~Yang, B.~Friesen, T.~Kurth, B.~Cook, and S.~Williams, ``{Toward Automated
  Application Profiling on Cray Systems},'' in \emph{Cray User Group Conference
  (CUG)}, 2018.

\bibitem{yang2020metho}
\BIBentryALTinterwordspacing
C.~Yang. {Hierarchical Roofline Analysis: How to Collect Data using Performance
  Tools on Intel CPUs and NVIDIA GPUs}. [Online]. Available:
  \url{https://arxiv.org/abs/2009.02449}
\BIBentrySTDinterwordspacing

\bibitem{madsen2020timemory}
J.~R. Madsen, M.~G. Awan, H.~Brunie, J.~Deslippe, R.~Gayatri, L.~Oliker,
  Y.~Wang, C.~Yang, and S.~Williams, ``{Timemory: Modular Performance Analysis
  for HPC},'' in \emph{International Conference on High Performance
  Computing}.\hskip 1em plus 0.5em minus 0.4em\relax Springer, 2020, pp.
  434--452.

\bibitem{yang2019hierarchical}
\BIBentryALTinterwordspacing
C.~Yang, T.~Kurth, and S.~Williams, ``{Hierarchical Roofline Analysis for GPUs:
  Accelerating Performance Optimization for the NERSC-9 Perlmutter System},''
  \emph{Concurrency and Computation: Practice and Experience}, p. e5547, 2019.
  [Online]. Available: \url{https://doi.org/10.1002/cpe.5547}
\BIBentrySTDinterwordspacing

\bibitem{yang2018empirical}
C.~Yang, R.~Gayatri, T.~Kurth, P.~Basu, Z.~Ronaghi, A.~Adetokunbo, B.~Friesen,
  B.~Cook, D.~Doerfler, L.~Oliker, J.~Deslippe, and S.~Williams, ``{An
  Empirical Roofline Methodology for Quantitatively Assessing Performance
  Portability},'' in \emph{2018 IEEE/ACM International Workshop on Performance,
  Portability and Productivity in HPC (P3HPC)}.\hskip 1em plus 0.5em minus
  0.4em\relax IEEE, 2018, pp. 14--23.

\bibitem{yang2020gpp}
\BIBentryALTinterwordspacing
C.~Yang. {8 Steps to 3.7 TFLOP/s on NVIDIA V100 GPU: Roofline Analysis and
  Other Tricks}. [Online]. Available: \url{https://arxiv.org/abs/2008.11326}
\BIBentrySTDinterwordspacing

\bibitem{ding2019instruction}
N.~Ding and S.~Williams, ``{An Instruction Roofline Model for GPUs},'' in
  \emph{2019 IEEE/ACM Performance Modeling, Benchmarking and Simulation of High
  Performance Computer Systems (PMBS)}.\hskip 1em plus 0.5em minus 0.4em\relax
  IEEE, 2019, pp. 7--18.

\bibitem{ibrahim2019performance}
K.~Z. Ibrahim, S.~Williams, and L.~Oliker, ``{Performance Analysis ff GPU
  Programming Models Using the Roofline Scaling Trajectories},'' in
  \emph{International Symposium on Benchmarking, Measuring and
  Optimization}.\hskip 1em plus 0.5em minus 0.4em\relax Springer, 2019, pp.
  3--19.

\bibitem{powerroofline}
J.~W. {Choi}, D.~{Bedard}, R.~{Fowler}, and R.~{Vuduc}, ``{A Roofline Model of
  Energy},'' in \emph{2013 IEEE 27th International Symposium on Parallel and
  Distributed Processing}, 2013, pp. 661--672.

\bibitem{alexpowerroofline}
A.~{Lopes}, F.~{Pratas}, L.~{Sousa}, and A.~{Ilic}, ``{Exploring GPU
  Performance, Power And Energy-Efficiency Bounds with Cache-aware Roofline
  Modeling},'' in \emph{2017 IEEE International Symposium on Performance
  Analysis of Systems and Software (ISPASS)}, 2017, pp. 259--268.

\bibitem{totocuda2020}
\BIBentryALTinterwordspacing
``Benchmarks and unit tests for gpus,'' accessed: 2020-08-01. [Online].
  Available:
  \url{https://github.com/PointKernel/toto-cuda/blob/master/benchmark/simpleOverhead.cu}
\BIBentrySTDinterwordspacing

\bibitem{ert2020}
\BIBentryALTinterwordspacing
``{Empirical Roofline Toolkit (ERT)},'' accessed: 2020-08-01. [Online].
  Available:
  \url{https://bitbucket.org/berkeleylab/cs-roofline-toolkit/src/master/}
\BIBentrySTDinterwordspacing

\bibitem{wang2020pmbs}
\BIBentryALTinterwordspacing
Y.~Wang, C.~Yang, S.~Farrel, T.~Kurth, and S.~Williams, ``{Hierarchical
  Roofline Performance Analysis for Deep Learning Applications},'' in
  \emph{2020 IEEE/ACM Performance Modeling, Benchmarking and Simulation of High
  Performance Computer Systems (PMBS)}. [Online]. Available:
  \url{https://arxiv.org/abs/2009.05257}
\BIBentrySTDinterwordspacing

\bibitem{roofline2020}
\BIBentryALTinterwordspacing
``{Roofline Methodology on NVIDIA GPUs}.'' [Online]. Available:
  \url{https://gitlab.com/NERSC/roofline-on-nvidia-gpus}
\BIBentrySTDinterwordspacing

\bibitem{gu2018recent}
J.~Gu, Z.~Wang, J.~Kuen, L.~Ma, A.~Shahroudy, B.~Shuai, T.~Liu, X.~Wang,
  G.~Wang, J.~Cai \emph{et~al.}, ``Recent advances in convolutional neural
  networks,'' \emph{Pattern Recognition}, vol.~77, pp. 354--377, 2018.

\bibitem{hahnloser2000digital}
R.~H. Hahnloser, R.~Sarpeshkar, M.~A. Mahowald, R.~J. Douglas, and H.~S. Seung,
  ``Digital selection and analogue amplification coexist in a cortex-inspired
  silicon circuit,'' \emph{Nature}, vol. 405, no. 6789, pp. 947--951, 2000.

\bibitem{okuta2017cupy}
R.~Okuta, Y.~Unno, D.~Nishino, S.~Hido, and C.~Loomis, ``Cupy: A
  numpy-compatible library for nvidia gpu calculations,'' in \emph{Proceedings
  of Workshop on Machine Learning Systems (LearningSys) in The Thirty-first
  Annual Conference on Neural Information Processing Systems (NIPS)}, 2017.

\bibitem{kernels2020}
\BIBentryALTinterwordspacing
``{Conv2D and LSTM Kernels and Results},'' accessed: 2020-08-01. [Online].
  Available: \url{https://github.com/PointKernel/tf-perf-kernels}
\BIBentrySTDinterwordspacing

\bibitem{guennebaud2010eigen}
G.~Guennebaud, B.~Jacob \emph{et~al.}, ``Eigen,'' \emph{URl: http://eigen.
  tuxfamily. org}, 2010.

\end{thebibliography}

\end{document}